\documentclass[fleqn,10pt]{SelfArx} % Document font size and equations flushed left

\usepackage{lipsum} % Required to insert dummy text. To be removed otherwise

\definecolor{color1}{RGB}{0,0,90} % Color of the article title and sections
\definecolor{color2}{RGB}{0,20,20} % Color of the boxes behind the abstract and headings

\usepackage[hyphens]{url}
\usepackage{hyperref} % Required for hyperlinks
\hypersetup{hidelinks,colorlinks,breaklinks=true,urlcolor=color2,citecolor=color1,linkcolor=color1,bookmarksopen=false,pdftitle={Title},pdfauthor={Author}}

\usepackage{adjustbox}
\usepackage{amssymb}
\usepackage{tabularx}
\usepackage{booktabs}  % for top-, mid- and bottomrule
\usepackage{makecell}
\usepackage[numbers, sort&compress]{natbib}
\usepackage{multirow}

\makeatletter
\newcommand\footnoteref[1]{\protected@xdef\@thefnmark{\ref{#1}}\@footnotemark}
\makeatother

\usepackage{float}

\JournalInfo{Draft Version} % Journal information
\Archive{ } % Additional notes (e.g. copyright, DOI, review/research article)

\PaperTitle{VANPY: Voice Analysis Framework} % Article title

\Authors{Gregory Koushnir\thanks{koushgre@post.bgu.ac.il}, Michael Fire\thanks{mickyfi@post.bgu.ac.il}, Galit Fuhrmann Alpert\thanks{fuhrmann@bgu.ac.il}, Dima Kagan\thanks{kagandi@post.bgu.ac.il}}
\affiliation{\textit{Department of Software and Information Systems Engineering, Ben-Gurion University of the Negev, Israel}}

 % Keywords - if you don't want any simply remove all the text between the curly brackets
\newcommand{\keywordname}{Keywords} % Defines the keywords heading name

\Abstract{

Voice data is increasingly being used in modern digital communications, yet there is still a lack of comprehensive tools for automated voice analysis and characterization. To this end, we developed the VANPY (Voice Analysis in Python)  framework for 
automated pre-processing, feature extraction, and classification of voice data. The VANPY is an open-source end-to-end comprehensive framework that was developed for the purpose of speaker characterization from voice data. The framework is designed with extensibility in mind, allowing for easy integration of new components and adaptation to various voice analysis applications. It currently incorporates over fifteen voice analysis components - including music/speech separation, voice activity detection, speaker embedding, vocal feature extraction, and various classification models. 

Four of the VANPY's components were developed in-house and integrated into the framework to extend its speaker characterization capabilities: gender classification, emotion classification, age regression, and height regression. The models demonstrate robust performance across various datasets, although not surpassing state-of-the-art performance.

As a proof of concept, we demonstrate the framework's ability to extract speaker characteristics on a use-case challenge of analyzing character voices from the movie "Pulp Fiction." The results illustrate the framework's capability to extract multiple speaker characteristics, including gender, age, height, emotion type, and emotion intensity measured across three dimensions: arousal, dominance, and valence. Quantitatively, our gender classifier achieved 100\% accuracy with no misses, height estimation resulted in a mean absolute error (MAE) of 5.5 cm for men and 10.2 cm for women, and age estimation yielded an MAE of 8.5 years for men and 9.9 years for women. Although emotion classification remains more challenging, our analysis discusses these results, indicating promising potential.
}

\begin{document}

\flushbottom % Makes all text pages the same height

\maketitle % Print the title and abstract box 
\section{Introduction} % The \section*{} command stops section numbering

\label{sec:int}

Voice data has become a key component in all media channels, such as phone calls, voice searches, and videos. 
Recent statistics published in 2024 indicate that the number of voice assistant users in the United States increased from approximately 132 million in 2021 to approximately 150 million in 2023, reflecting a steady increase in hands-free interaction~\cite{demandsage_2024}. Further data show that 71\% of smart speakers consumers prefer to perform voice searches rather than typing, illustrating a strong shift toward voice-based technologies~\cite{PricewaterhouseCoopers, businessdasher_2024}.

As a digital evolution of traditional radio broadcasting, podcasts have revolutionized voice media by offering personalized, on-demand audio content across diverse topics and formats. 
According to `The Podcast Consumer 2024' report, 67\% of the population aged 12 and above have listened to a podcast at least once, with 47\% being monthly listeners and 34\% tuning in weekly~\cite{Edison_Research_2024}.

While vast amounts of voice data are accumulating, there is still a lack of tools that could be used to process the data and extract valuable information automatically. 
Although voice analysis and characterization are relatively popular areas for computational research and applications~\cite{schultz2007speaker, anguera2012speaker, NAGRANI2020101027}, they continue to lag behind the extensively researched field of computer vision in terms of the number of research publications~\cite{alam2020survey} and its demonstrated robust large-scale data processing capabilities.

However, speech analysis has the potential to yield valuable and unique insights.
Potentially, it could be possible to draw inferences about the style and personality of speakers, as well as their psychological and physical states, using audio data only ~\cite{mokhtari2008speaking}. Characteristics of a speaker may include gender classification~\cite{gupta2016support}, age estimation~\cite{qawaqneh2017dnn}, accent detection~\cite{parikh2020english}, emotional state classification~\cite{jason2020appraisal}, vocal health (e.g., detection of vocal strain or damage, such as hoarseness, breathiness, or vocal fatigue)~\cite{hegde2019survey}, and speech style (e.g., use of pauses and speech rate)~\cite{amano2009classifying}.

A robust and reliable speaker characterization framework is essential for many practical applications. By extracting physiological and psychological aspects of speaker characteristics from a voice recording, a speaker characterization framework can provide valuable information about the speaker's identity and traits, which can be used for various purposes such as speech crime investigation, health monitoring, human-computer interaction, and more~\cite{Kinnunen}. 

Some fundamental tasks that benefit from voice analysis include - figuring out if there is a speaker within an audio segment (\textit{detection/segmentation}), identifying the voice owner (\textit{identification} or \textit{classification} for a closed set of speakers), authenticating someone's voice (\textit{verification}), identifying and partitioning the same speaker through the audio stream (\textit{diarization}), retrieving static and dynamic labeled characteristics (\textit{characterization})~\cite{schultz2007speaker, jason2020appraisal}. 

Complex tasks and applications manipulate voice-based data insights. They can be used to address a variety of challenges, for example, whether a speaker suffers from voice pathology~\cite{reid2022development}, depression~\cite{6359792}, fear and nervousness~\cite{kumar2016efficient}, Parkinson's disease~\cite{8701961}, what is their smoking status~\cite{Smoking}.

Ongoing advancements in machine learning (ML) and deep learning (DL) have significantly enhanced the capabilities of analyzing voice data, enabling robust feature extraction, classification, and regression tasks. Voice data can be processed using techniques like feature extraction, speaker embeddings, and time/frequency domain representations, which provide detailed insights into speaker characteristics.

Feature extraction methods such as Mel Frequency Cepstral Coefficients (MFCCs)~\cite{tiwari2010mfcc}, Zero-Crossing-Rate (ZCR) \cite{ganchev2005comparative}, and Spectral Centroid~\cite{tiwari2010mfcc} capture acoustic and temporal properties of speech. %These features are pivotal for downstream tasks such as emotion recognition and speaker identification. 
Additionally, speaker embeddings like x-vectors~\cite{snyder2018x} and d-vectors~\cite{variani2014deep} leverage neural networks to encode latent speaker characteristics, enhancing the robustness of classification models.

Classification and regression tasks are pivotal for speaker characterization, enabling robust analysis of speaker attributes such as gender, age, height, and emotional states~\cite{gupta2016support, kaushik2021end, zeng2019spectrogram, wagner2022dawn}. Additionally, speech-to-text (STT) systems like OpenAI Whisper~\cite{https://doi.org/10.48550/arxiv.2212.04356} allow for further text-based analyses, including topic classification and sentiment evaluation~\cite{lewis2019bart}.
Recent advancements in large language models (LLMs) have further expanded the scope of voice technologies, enhancing their ability to provide natural and efficient interactions. For instance, integration of LLMs into voice technologies opened new avenues for personalized and context-aware audio content delivery~\cite{flavian2023effects, akdim2023perceived, doi:10.1080/08874417.2024.2312858}.

While voice analysis research encompasses multiple challenges, we focus on speaker characterization from auditory data in this paper. We introduce the Voice Analysis in Python (VANPY)\footnote{The code is available at \url{https://github.com/griko/vanpy}} framework to address this. This end-to-end open-source framework allows the building and execution a user-defined modular pipeline to allow automatic preprocessing, feature extraction, and classification components for text-independent speaker characterization. 
%A comparison of our framework with others is shown in Table~\ref{tab:capabilities_comparison}. 
The framework is built generically and is intended to be applicable to a wide range of voice analysis applications. More than fifteen components are implemented in the framework, and it is designed in a modular fashion that could easily be extended with new ones. The currently incorporated components include music/speech separation, Voice Activity Detection (VAD), speaker embedding, vocal feature extraction, gender identification, emotion classification, height and age estimation, and other components. Some of those components are third-party-based state-of-the-art models, while other components were trained in-house to achieve near state-of-the-art performance. 

We demonstrate the robust performance of VANPY on multiple benchmark speech datasets, including VoxCeleb2 multilingual dataset~\cite{Chung18b}, The DARPA TIMIT Acoustic-Phonetic Continuous Speech Corpus (TIMIT)~\cite{discacoustic}, and Mozilla Common Voice v10.0 English dataset~\cite{mozilla_2020}, as will be shown in the following sections.

\begin{figure*}
 \centering
 \captionsetup{justification=centering}
 \includegraphics[width=\linewidth]{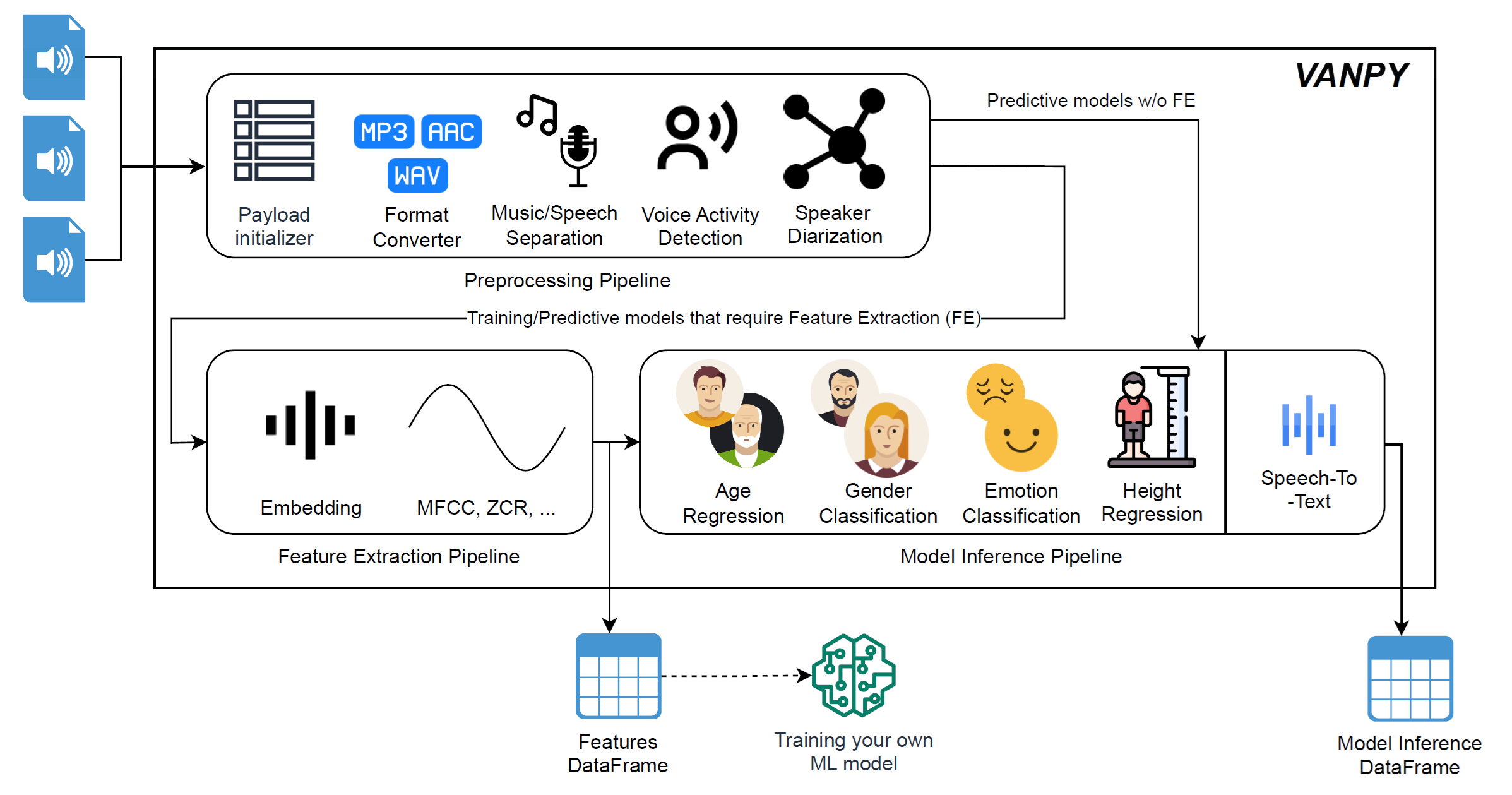}
  \caption{\textit{VANPY's} architecture.}
	\label{figure:architecture}
\end{figure*} 

The remainder of this article is organized as follows: 
Section~\ref{sec:rw} reviews the relevant literature across audio preprocessing, feature extraction, and speaker characterization models. In Section~\ref{sec:methods}, we discuss our implementation of the voice characterization pipeline, which we refer to as the VANPY framework, and describe the machine learning models trained to enhance VANPY's functionality for speaker characteristics. In Section~\ref{sec:results}, we present comprehensive results of our gender classifier, emotion recognition classifier, age regressor, and height regressor and provide a performance summary of our custom-trained models along with a use-case example. Section~\ref{sec:discussion} discusses our findings and insights. Finally, Section~\ref{sec:conclusions} summarizes our work and presents concluding remarks. 

\section{Related Work}
\label{sec:rw}
The field of speaker characterization encompasses the analysis of physiological and psychological speaker traits, such as gender, age, height, accent, and expressed emotion~\cite{phdthesis_Safavi}. Techniques for speaker characterization typically rely on machine learning models to infer speaker attributes from extracted voice features. These approaches are complemented by preprocessing methods, such as voice activity detection and feature extraction, to prepare the data for downstream analysis.

Table~\ref{tab:capabilities_comparison} provides an overview of speaker characterization approaches and frameworks. The first section presents individual studies focusing on specific characterization tasks.
The second section of the table highlights general-purpose speech-processing frameworks that provide preprocessing and feature extraction capabilities. These frameworks, while offering essential functionalities like VAD and diarization, typically lack comprehensive speaker characterization capabilities. In contrast, our VANPY framework bridges this gap by integrating both fundamental preprocessing components and advanced speaker characterization models.

\begin{table*}[ht]
\centering
\footnotesize
\begin{adjustbox}{width=1\linewidth}
\begin{tabular}{lp{30mm}p{45mm}*{5}{c}p{70mm}}
\toprule
\textbf{Framework/Tool} & \textbf{\makecell[l]{Audio Preprocessing}} & \textbf{\makecell[l]{Feature Extraction and \\ Speaker Representations}} & \multicolumn{5}{c}{\textbf{Speaker Characterization}} & \textbf{\makecell[l]{Other Capabilities and \\ incorporated ML models}} \\ 
\cmidrule(lr){4-8}
& & & \rotatebox{90}{\textbf{Gender ID}} & \rotatebox{90}{\textbf{Dialect/Accent ID}} & \rotatebox{90}{\textbf{Emotion Classification}} & \rotatebox{90}{\textbf{Height Estimation}} & \rotatebox{90}{\textbf{Age Estimation}} & \\ 
\midrule
\citeauthor{shafran2003voice}~(\citeyear{shafran2003voice}) & - & - & \checkmark & 7 & 6 & - & 5 & - \\ 
AGENDER ~(\citeyear{muller2006automatic}) & - & - & \checkmark & - & - & - & 4 & - \\ 
\citeauthor{schultz2007speaker} et al.~(\citeyear{schultz2007speaker}) & - & - & \checkmark & 2 & - & - & - & Speaker Verification, Language ID, Proficiency Level, Attentional State speaker \\ 
\citeauthor{5670700}~(\citeyear{5670700}) & - & - & \checkmark & 3 & - & - & 3 & - \\ 
\citeauthor{gupta2016support}~(\citeyear{gupta2016support}) & - & - & \checkmark & - & - & - & - & - \\ 
\citeauthor{qawaqneh2017dnn}~(\citeyear{qawaqneh2017dnn}) & - & - & \checkmark & - & - & - & 4 & - \\ 
\citeauthor{jahangir2020text}~(\citeyear{jahangir2020text}) & - & - & \checkmark & - & - & - & - & Speaker Verification \\ 
\citeauthor{kaushik2021end}~(\citeyear{kaushik2021end}) & - & - & - & - & - & \checkmark & \checkmark & - \\
\citeauthor{lastow2022language}~(\citeyear{lastow2022language}) & - & - & \checkmark & - & - & - & \checkmark & - \\ 
\midrule
pyAudioAnalysis~(\citeyear{giannakopoulos2015pyaudioanalysis}) & Silence removal, Speaker diarization & MFCCs. ZCR, Chroma Vector, Chroma Deviation, Energy, Entropy of Energy, Spectral Centroid, Spectral Spread, Spectral Entropy, Spectral Flux, and Spectral Rolloff & - & - & - & - & - & Fixed and HMM-based segmentation and classification, Train and use supervised models for classification and regression, Audio thumbnailing, Feature Visualization, Batch format conversion, Audio recording \\ 
ESPnet (2018) & CTC segmentation & FFT, Speaker ID embedding, Language ID embedding & - & - & - & - & - & Automatic Speech Recognition, Text-To-Speech, Speech Enhancement, Speech Translation \& Machine Translation, Voice conversion, Spoken Language Understanding, Speech Summarization, Singing Voice Synthesis \\ 
\midrule
VANPY (2024) & INA speech-music detection, Pyannote Segmentation, Pyannote Diarization, Silero VAD & MFCC, ZCR, Tonnetz, SpeechBrain Speaker ID embedding, Pyannote Speaker ID x-vector embedding, Spectral Centroid, Spectral Bandwidth, Spectral Contrast, Spectral Flatness, Fundamental Frequencies & \textcolor{blue}{\checkmark} & - & \textcolor{blue}{7, 8} & \textcolor{blue}{\checkmark} & \textcolor{blue}{\checkmark} & SpeechBrain emotion recognition, Audio classification (YAMNet), Facebook Wav2vec 2.0 arousal-dominance-valence estimation, Language ID (OpenAI Whisper), Speech-To-Text (Facebook Wav2Vec 2.0, OpenAI Whisper), Speech Enhancement (MetricGAN+, SepFormer) \\ 
\bottomrule
\end{tabular}
\end{adjustbox}
\caption{Comparison of capabilities of various speech and audio processing frameworks. Number instead of \(\checkmark\) represents the number of classes for the classification variant of the capability. The blue color indicates models trained as a part of this research.}
\label{tab:capabilities_comparison}
\end{table*}

\subsection{Audio Preprocessing}

Preprocessing is crucial for consistent, high-quality input in voice analysis. Basic operations like resampling and amplitude normalization align sampling rates and mitigate recording differences. These steps often precede more advanced techniques such as diarization and VAD~\cite{848229}, which isolate speech segments. Recent advances in deep learning have significantly improved the performance of these preprocessing techniques.

\citet{Bredin2021} proposed the Pyannote segmentation model, which demonstrates robust performance across benchmark datasets by integrating VAD with speaker diarization. Silero VAD~\cite{Silero} supports over 100 languages and has been shown to outperform other VAD models like Picovoice\footnote{https://picovoice.ai/} and WebRTC\footnote{https://webrtc.org/} in precision-recall metrics. INA’s music-speech detection model~\cite{doukhan2018ina} excels at distinguishing speech from music and achieved first place in the MIREX2018 speech detection challenge.\footnote{https://music-ir.org/mirex/wiki/MIREX\_HOME} These preprocessing techniques ensure clean and reliable audio input for downstream speaker characterization.

Several open-source frameworks, such as Kaldi~\cite{Povey_ASRU2011} and pyAudioAnalysis~\cite{giannakopoulos2015pyaudioanalysis}, also integrate these essential preprocessing components, including resampling and VAD, thereby ensuring consistent audio input for subsequent speaker
characterization.

\subsection{Feature Extraction and Speaker Representations}

Voice data can be represented and analyzed using various techniques, including feature extraction, time/frequency domain representations, and speaker embeddings. These methods provide unique perspectives on the acoustic and temporal properties of speech, enabling robust speaker characterization and identification.

Feature extraction from voice segments has seen several breakthroughs in recent decades~\cite{8882461, prabakaran2019review, li2019vocal, al2019new, TIRUMALA2017250}. Known features include Mel Frequency Cepstral Coefficients (MFCC)~\cite{tiwari2010mfcc, ganchev2005comparative, nakagawa2011speaker}, which capture spectral envelopes approximating human auditory perception, Shifted Delta Cepstral coefficients (SDC)~\cite{torres2002approaches}, which capture temporal dynamics across multiple frames, and Zero-Crossing-Rate (ZCR)~\cite{khan2012hindi, ramaiah2016multi}, measuring frequency content changes through zero-amplitude axis crossings. Frequency-based features include fundamental frequency (F0) for pitch perception, jitter for pitch perturbations, and formant frequencies characterizing vocal tract resonance. Energy and amplitude parameters like shimmer quantify amplitude variations, while harmonics-to-noise ratio (HNR) compares harmonic content to noise. Additional spectral and temporal features such as loudness peaks and voiced region rates provide further insights~\cite{eyben2015geneva}. Modern open-source libraries like \textit{librosa}~\cite{brian_mcfee_2022_6097378} facilitate extraction of spectral features, including MFCC, spectral centroid, bandwidth, contrast, flatness, and roll-off. The library also provides methods of feature manipulation capabilities like delta features for capturing temporal dynamics. 

Many models process inputs as raw waveforms (time domain) or spectrograms (frequency domain) arrays or images~\cite{jia2021speaker, zhang2018text, nagrani2017voxceleb, palaz2015analysis, muckenhirn2018towards, 8639585}, offering an alternative perspective that capitalizes on deep learning architectures designed for image-like data.

Over the past decade, speaker embeddings have emerged as powerful neural representations that capture the unique characteristics of a speaker's voice in a compact, fixed--
dimensional vector space. These advanced models have demonstrated superior performance in speaker identification tasks \cite{BAI202165}, surpassing traditional methods that directly utilize raw acoustic features. These models produce speaker embeddings that encapsulate deeper properties of a speaker's voice. Embeddings for speaker characterization can be generally applied to text-independent tasks, such as i-vectors~\cite{dehak2010front, kanagasundaram2011vector}, d-vectors~\cite{variani2014deep,jung2017d,8462628}, and x-vectors~\cite{snyder2017deep, zeinali2018convolutional}. In particular, x-vectors are learned via time-delay neural networks (TDNN) to encode speaker-specific information, and \textit{PyAnnote~2.0}~\cite{Bredin2021} provides an implementation optimized for speaker diarization. 
More recent studies have enhanced these architectures, with ECAPA-TDNN~\cite{Desplanques_2020} 
expanding the x-vector approach through channel attention and advanced aggregation mechanisms. This method has been integrated into \textit{SpeechBrain}~\cite{speechbrain, Dawalatabad_2021}, demonstrating state-of-the-art performance and broad applicability in speaker-related applications.

\subsection{Speaker Characterization Models}
\label{ssec:model_inf}

Various studies have attempted to address the issue of speaker characterization over the years, targeting attributes such as gender, age, dialect, height, and emotion. Early work by~\citet{shafran2003voice} combined multiple classifications (gender, age, dialect, emotion) in one system using Hidden Markov Models (HMMs) with spectral and prosodic features. While achieving high accuracy for gender classification (96\%), performance varied significantly across other characteristics. Their dataset, collected from a customer-care system, suffered from class imbalance, yielding lower accuracies: 50\% for seven-class dialect, 69\% for seven-class emotion, and 70\% for three-group age classification. Notably, models trained to predict multiple characteristics outperformed single-task models. The addition of pitch features to MFCC consistently improved performance across all models.
In 2006,~\citet{muller2006automatic} demonstrated that different acoustic features contribute varying importance to gender-age classification. Their system incorporated acoustic features like F0 and HNR, alongside temporal features including articulation rate (syllables per second) and pause-to-utterance ratios. Among the five classification methods evaluated, their Artificial Neural Network (ANN) achieved the highest accuracy: 93.14\% for binary gender classification and 65.5\% for the more challenging eight-class combined gender-age classification.

Research by~\citet{5670700} demonstrated the superiority of Support Vector Machine (SVM) over vector quantization and Gaussian mixture modeling for speaker characterization. Using a dataset of Australian English speakers with balanced demographics and controlled recording conditions, and extracting a set of acoustic features, their system achieved 100\% accuracy for gender classification, 98.8\% for three-class age groups, and 98.7\% for three-class accent detection. Later,~\citet{gupta2016support} demonstrated similar performance using SVM for gender classification, achieving 99.5\% accuracy on a multilingual Indian speech corpus.
In 2017,~\citet{qawaqneh2017dnn} investigated joint age-gender classification using Deep Neural Network (DNN). They enhanced the traditional MFCC-based approach by incorporating SDC features. Testing on the Age-Annotated Database of German Telephone Speech, which contains seven age-gender categories, their system achieved 57.2\% accuracy. The telephone channel's bandwidth limitations and potential noise likely presented additional challenges beyond those inherent in training complex models with limited data.
\citet{jahangir2020text} later demonstrated the superiority of Deep Neural Networks over traditional classification methods using the LibriSpeech audiobook dataset. Their approach enhanced MFCC features with custom time-domain characteristics, improving accuracy by 25\% to achieve 92.9\% for gender classification. 

In 2021,~\citet{kaushik2021end} demonstrated a combination of the attention mechanism with Long Short Term Memory (LSTM) encoder for joint age-height regression prediction from speech segments in the TIMIT. Their model achieved state-of-the-art performance for age estimation with a Mean Absolute Error (MAE) of 5.62/6.08 years for male and female speakers, respectively. For height estimation, the MAE was 5.24/5.09 centimeters for male and female speakers, respectively.

More recently,~\citet{lastow2022language} advanced the field using embeddings from self-supervised pre-trained models. By fine-tuning these speech representations, their models achieved state-of-the-art results on the TIMIT dataset: 99.8\% accuracy for gender classification and remarkably low age  MAE of 4.11/4.44 years for male/female speakers. Their work demonstrated the effectiveness of transfer learning from large-scale pre-trained models in addressing data efficiency challenges for speaker characteristic prediction.

Emotion classification and regression have also been extensively studied.~\citet{zeng2019spectrogram} proposed a spectrogram-based model leveraging both shallow and deep CNNs for multi-task emotion classification, achieving 66\% accuracy in an 8-class emotion recognition task on the RAVDESS dataset, while~\citet{wagner2022dawn} developed transformer-based models for dimensional emotion analysis along arousal, dominance, and valence axes, demonstrating strong performance with a concordance correlation coefficient (CCC) of 0.638 on the MSP-Podcast corpus.

Recently, ~\citet{burkhardt2023speechbasedagegenderprediction} leveraged wav2vec 2.0 for speaker characterization, achieving 100\% accuracy for gender classification and 7.1 years MAE for age estimation on TIMIT. These advancements highlight the continued progress in robust speaker characterization models.

\section{Methods}
\label{sec:methods}
\subsection{The VANPY Framework}
\label{sec:framework}

We developed the VANPY framework to address the task of speech characterization as a sequence of interconnected stages, collectively forming a comprehensive pipeline: (1)\textit{ audio file preprocessing,} (2) \textit{feature extraction}, and (3) \textit{model interference}. The proposed VANPY framework ensures a simple configuration and execution of these integral stages. Each sub-pipeline within this framework comprises multiple components executed sequentially, employing multi-threaded operations wherever applicable for optimized performance. These components utilize an Inter-Component Payload (referred to as ``\textit{payload}'') as an input and output object. We implemented the payload as a Pandas DataFrame object~\cite{reback2020pandas}, accompanied by corresponding metadata. While the DataFrame stores the data, the metadata houses additional details and column names such as the actual paths column, lists of previously processed paths, performance measurements, features, and model inference, all of which can facilitate data filtering.

\begin{figure*}
 \centering
 \captionsetup{justification=centering}
 \includegraphics[width=\linewidth]{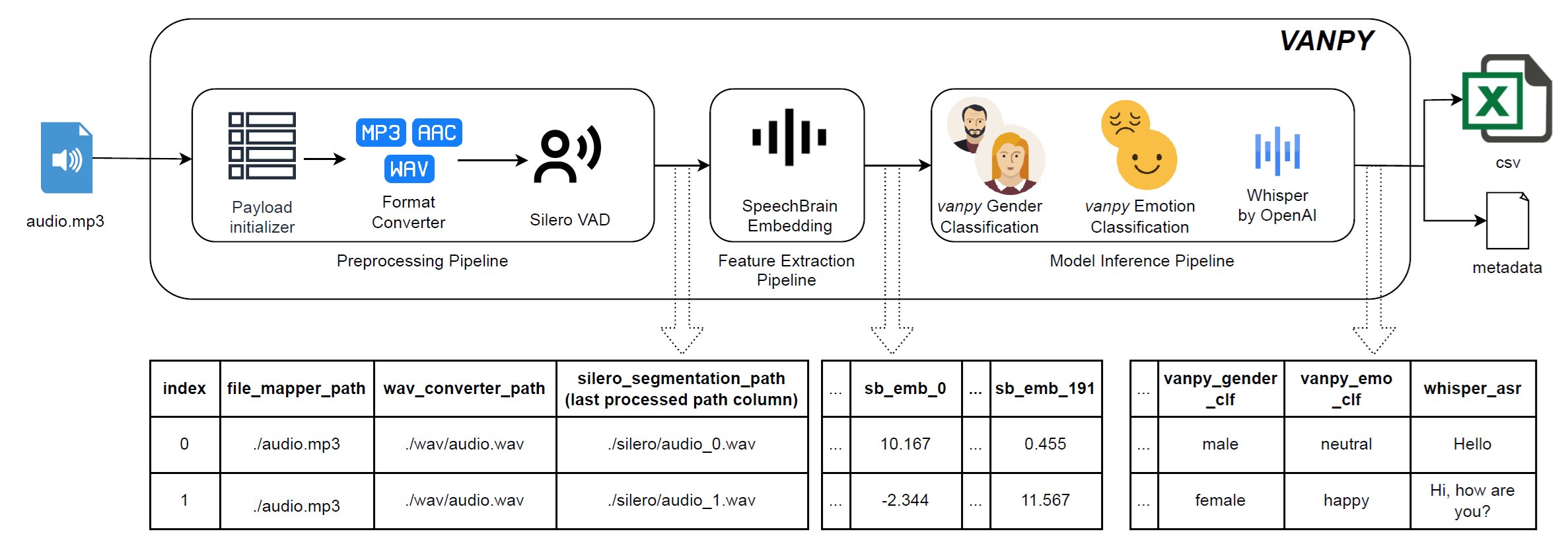}
  \caption{VANPY usage example}
	\label{figure:example}
\end{figure*} 
  
Figure~\ref{figure:architecture} provides an overview of VANPY's general architecture. Based on the specific task, various component combinations may be used. Subsequently, we will outline each pipeline and explain its purposes and the components used. Additionally, we give a real-world inspired example to demonstrate the process of extracting gender, emotion, and transcription of each speaker from a given audio file. The flow of the framework analysis on the example is illustrated in Figure~\ref{figure:example}.

\subsubsection{Preprocessing Pipeline}
\label{sec:preprocess}
The framework's preprocessing pipeline deals with audio files' format conversion and voice segment cutting. The format conversion includes the transformations of a number of channels, as well as sample and bitrate frequencies if required. For the voice segment cuttings and neglecting of low-amplitude parts, it is common to use VAD algorithms~\cite{848229}.

\paragraph{Implemented Preprocessing Components}
Our preprocessing pipeline incorporates the following components, each designed to handle a specific element of audio processing: 

\begin{itemize}
    \item \textit{Payload Initializer} - generates an initial payload by listing audio files in a folder or loading previously saved payload.
    \item \textit{WAV converter} - converts audio files' format to WAV format using FFMPEG~\cite{tomar2006converting}.
    \item \textit{INA's music-speech detection}~\cite{doukhan2018ina} - separates music and speech parts.
    \item \textit{Pyannote VAD}~\cite{Bredin2021} - VAD component.
    \item \textit{Silero VAD}~\cite{Silero} - VAD component.
    \item \textit{Pyannote Speaker Diarization}~\cite{bredin2020pyannote} - a speaker diarization tool that segments and clusters audio data by speaker identity, incorporating a VAD unit within its pipeline. It achieves Diarization Error Rates (DER) ranging from 8.17\% to 64\% across different datasets
    \item \textit{MetricGAN+}~\cite{fu2021metricgan} speech enhancement model, which improves speech quality, optimized on the perceptual evaluation of speech quality (PESQ) score. PESQ score 3.15 on the \textit{VoiceBank-DEMAND} dataset.
    \item \textit{SepFormer}~\cite{subakan2021attention} speech enhancement model. PESQ score 2.2 on the \textit{WHAM!} dataset.
\end{itemize}

\paragraph{Preprocessing Example.} At the beginning of VANPY execution, consider an "audio.mp3" file featuring two speakers of different genders exchanging greetings. The execution commences with the preprocessing pipeline. The first component, termed as \textit{Payload Initializer}, systematically maps the audio files located in a specified directory, subsequently creating a payload. A single row is present within the payload's DataFrame, detailing the path to the input file under the "file\_mapper\_path" column. Correspondingly, the actual paths column within the payload's metadata is assigned as "file\_mapper\_path."

As the payload advances to the ensuing components, the \textit{WAV Converter} comes into operation, transforming the "audio.mp3" to "audio.wav." This conversion is executed with a reduced single channel, a sampling frequency of 16kHz, and a 256 kb/second bitrate. This process contributes an additional "wav\_converter\_path" column to the DataFrame, and it is designated as the new actual paths column within the metadata.

Subsequently, the \textit{Silero VAD} component is employed to identify two segments with active voice. This process leads to the segmentation of "audio.wav" into two separate files, "audio\_0.wav" and "audio\_1.wav," based on the start and end time markers as suggested by the VAD component. The resulting DataFrame object encompasses two rows, with the "silero\_segmentation\_path" as the newly appointed actual paths column. Then, this newly constructed payload is seamlessly transferred to the subsequent Feature Extraction pipeline.

\subsubsection{Feature Extraction Pipeline}
\label{sec:featureext}
The feature extraction process in voice analysis plays a pivotal role, especially in the context of machine learning models. In VANPY, this process involves transforming raw audio data into a structured and informative format that is understandable from a machine-learning perspective. A key distinction of the VANPY framework is its focus on speaker characterization, concentrating more on the speaker's attributes rather than the speech's content. VANPY primarily implements a subset of speaker embedding and acoustic parameter extraction methods. 

\paragraph{Implemented Feature Extraction Components.}
Below, we enumerate the components utilized for the extraction of feature and speaker-oriented latent vectors:
\begin{itemize}
    \item \textit{Librosa Feature Extractor} - extracts MFCC and its derivatives (delta-MFCC), ZCR, Spectral Centroid, Spectral Bandwidth, Spectral Contrast, Spectral Flatness, F0, and tonnetz by utilizing Librosa library~\cite{brian_mcfee_2022_6097378}.
    \item \textit{Pyannote Audio 2.0 Embedding}~\cite{bredin2020pyannote} - an x-vector speaker embedding implementation that neglects the Probabilistic Linear Discriminant Analysis part.
    \item \textit{SpeechBrain Embedding}~\cite{speechbrain} - an x-vector speaker embedding extraction with TDNN.
\end{itemize}

\paragraph{Feature Extraction Example.}
The Feature Extraction pipeline consists of a \textit{SpeechBrain Embedding} component. This component enriches the payload by appending a 192-dimensional SpeechBrain embedding for each voice-active segment previously extracted. This enhancement manifests as adding of 192 columns to the payload's DataFrame, specifically named "sb\_emb\_0" through "sb\_emb\_191." These new columns are duly registered in the metadata under the category of "feature columns." Subsequently, this augmented payload is transitioned to the Model Inference pipeline.

\subsubsection{Model Inference Pipeline}
\label{sec:clfpipeline}
Given a feature set that represents a voice segment, speaker characterization may be applied. It includes classification and regression models that attempt to acquire a physiological and psychological description of the speaker. Known classification and regression characterization models include gender~\cite{5670700, muller2006automatic, schultz2007speaker, gupta2016support, qawaqneh2017dnn, jahangir2020text}, age~\cite{5670700, muller2006automatic, qawaqneh2017dnn}, expressed emotion~\cite{speechbrain_2021}, height estimation~\cite{kaushik2021end}, and language and accent detection~\cite{schultz2007speaker, dehak2011language}.

The model inference pipeline of our framework executes classification/estimation/Speech-To-Text(STT)/ clustering models. 
Some classification components might be used on the audio files themselves, while others require a specific feature set they were trained on.

\paragraph{Implemented Model Inference Components}
Third-party pre-trained models predominantly utilize raw audio files as input for various tasks such as emotion recognition and STT conversion. Below is presented a list of third-party pre-trained models utilized in the framework:
    \begin{itemize}
        \item \textit{SpeechBrain}~\cite{speechbrain}  emotion recognition - emotion classifier trained on Interactive Emotional Dyadic Motion Capture (\textit{IEMOCAP}~\cite{Busso2008IEMOCAPIE}) dataset. It achieved 75.3\% average accuracy in four classes~\cite{speechbrain_2021}: neutral, angry, happy, and sad.
        \item \textit{Facebook Wav2Vec 2.0}~\cite{https://doi.org/10.48550/arxiv.2006.11477} STT - A model that generates speech embeddings from raw audio. These embeddings were pre-trained and fine-tuned for the STT task on 960 hours of Librispeech~\cite{panayotov2015librispeech}, 16kHz sampled speech audio. Word Error Rate (WER) 1.8/3.3\% on the clean/other test sets of the \textit{Librispeech} dataset.
        \item \textit{OpenAI Whisper}~\cite{https://doi.org/10.48550/arxiv.2212.04356} - general-purpose speech recognition model capable of detecting spoken language and retrieving state-of-the-art transcriptions from multiple languages.
        \item \textit{YAMNet}~\cite{Plakal_Ellis} classifier - classifies 521 audio event classes, trained on AudioSet-YouTube corpus. The model scores 0.306 for balanced mean average precision (mAP), and the balanced label-weighted label-ranking average precision (lwlrap) is 0.393 on the AudioSet-YouTube eval set.
        \item \textit{Wav2vec 2.0 Arousal, Dominance, and Valence}~\cite{wagner2022dawn} - prediction transformer-based model for dimensional emotion recognition that achieves state-of-the-art valence recognition performance on MSP-Podcast without using explicit linguistic information.
        \item \textit{Clustering Models} - given a list of features and classification and regression targets - can be applied to distinguish speakers  (assuming one speaker in any given segment) and speaker labeling purposes:  
        \begin{enumerate}
            \item \textit{Agglomerative Clustering} - as implemented in scikit-learn~\cite{scikit-learn}, can be utilized for speaker clustering based on the feature vectors. This hierarchical clustering method merges clusters based on the closest distance criterion.
            \item \textit{Gaussian Mixture Model (GMM) Clustering} - uses Gaussian Mixture Models to probabilistically assign segments to clusters based on the underlying distribution of the data. This method also uses the scikit-learn implementation.
            \item \textit{Cosine Distance Clustering (self-implemented, a private case of Agglomerative Clustering)} - implements Disjoint-Set clustering using a minimal distance threshold. This method receives an arbitrary feature set and performs clustering using the union-find algorithm on the disjoint set with respect to the given threshold. It assumes that for any unlabeled feature-vector X, if the cosine similarity distance to any labeled feature-vector Y is greater than the threshold, X should get the same label as Y. If vectors in different labeled clusters satisfy this condition, the clusters are merged and will have the same label.
        \end{enumerate}
    \end{itemize}

In contrast to the third-party models, the models trained as part of this work primarily use extracted features stored in the payload, obtained from the preceding feature extraction pipeline, for tasks including gender and emotion classification, and age and height regression. Models trained as part of this work are detailed in Section~\ref{sec:classifiers}.

\paragraph{Model Inference Example.}
In the example, the Model Inference pipeline contains a \textit{Gender Classifier}, \textit{Emotion Classifier}, and \textit{OpenAI Whisper} components. The Gender and Emotion Classifiers utilize the SpeechBrain embedding columns, which were previously extracted by the Feature Extraction pipeline and are currently present in the payload. These classification components operate on every row of the payload's DataFrame, appending "VANPY\_gender\_clf" and "VANPY\_emo\_clf" 
columns to the DataFrame and noting these in the metadata under model inference columns.

The \textit{OpenAI Whisper} component contributes an additional column, "whisper\_asr," populated with automatic speech recognition text for files "audio\_0.wav" and "audio\_1.wav." Upon completion of these steps, the payload's content will be ready to be saved. The resultant DataFrame offers the flexibility for subsequent filtering, permitting the inclusion or exclusion of preprocessing columns, feature columns, and model inference columns as per the research requirements.

\subsection{Incorporated models for Inferring Speaker Characteristics}
\label{sec:classifiers}
This section describes the models specifically built to extend the framework's capabilities. These models are emphasized with blue in Table~\ref{tab:capabilities_comparison}. The preprocessing and feature extraction for the models' training were done by utilizing VANPY.

\subsubsection{Gender Classifier}
\label{sssec:gender_clf}
\paragraph{Training and evaluating on VoxCeleb2 dataset.}
\label{sssec:gender_voxceleb}
The gender classifier was trained on the VoxCeleb2 dataset, which contains records of YouTube interviews of celebrities. A total of 4,196 speakers participated, each contributing to a range of 5 to 81 different interviews. The dataset was divided into training (845 females and 846 males), validation (396 females and 389 males), and testing (828 females and 819 males) sets; no speakers were shared between the sets. The test-set IDs are identical to those used in~\cite{DBLP:journals/corr/abs-2109-13510}. For our validation set, we randomly selected 800 IDs from their original train set, with the remaining 1691 IDs constituting our train set.

We have sampled a random audio track part for each video ID, leaving 135,512 samples.
The file format was converted from \textit{mp3} to \textit{wav}, 1-channel, with a sampling rate of 16 KHz. To neglect the low-energy parts, we applied the SileroVAD model, leaving only the first high-energy segment. To make a gender classification, we tested different feature sets and models.  

In our research, we explored four distinct feature sets to evaluate their effectiveness in gender classification tasks. The feature sets we employed are as follows: SpeechBrain x-vector embedding with 512 dimensions, SpeechBrain ECAPA embedding with 192 dimensions, Pyannote x-vector embedding with 512 dimensions (using a sliding window duration of 3.0 seconds and a step of 1.0 second), and a set of 31 features extracted using the Librosa library. Additionally, we experimented with a combination of SpeechBrain ECAPA embedding and Librosa features, resulting in a 223-dimensional feature set.

We evaluated XGboost~\cite{Chen:2016:XST:2939672.2939785}, SVM~\cite{cortes1995support}, Logistic Regression~\cite{cox1958regression}, and various architectures of Feed Forward Artificial Neural Network (ANN). For hyperparameter tuning, we employed the Optuna library~\cite{akiba2019optunanextgenerationhyperparameteroptimization}, which uses a validation set to optimize model performance. We conducted 200 study trials for XGBoost, Logistic Regression, and ANN models. Due to computational constraints, we limited the SVM to 100 trials. 

\paragraph{Evaluating on Mozilla Common Voice v10.0 English test set.}
\label{sssec:gender_mozilla}
We evaluated the models' effectiveness using a test set taken from the Mozilla Common Voice v10.0 English dataset. 
The test set was filtered to contain records with positive \textit{up\textunderscore votes} and zero \textit{down\textunderscore votes}, leaving 595 records. 
Pre-processing and evaluation were done in the same manner as for the VoxCeleb2 dataset. 

\paragraph{Evaluating on DARPA-TIMIT test set.}
\label{sssec:gender_timit}
We evaluated our models on the TIMIT test set, which offers a diverse range of American English dialects. This test set comprises 1,680 recordings from 168 different speakers (112 males and 56 females).
The pre-processing and evaluation procedures were consistent with those applied to the VoxCeleb2 dataset, with one notable exception: we omitted the VAD step. This omission was justified by the nature of the TIMIT recordings, which are precisely segmented and have a maximum duration of 7 seconds.

\subsubsection{Emotion Recognition Classifier}
\label{sssec:emo_clf}
We trained the emotion recognition classifier on the RAVDESS \cite{livingstone_steven_r_2018_1188976} dataset, which provides 1440 recorded audio fragments: 24 actors, 60 trials per actor across eight emotion classes (neutral, calm, happy, sad, angry, fearful, disgust, surprised) at two intensity levels (except the "natural" class). Our preprocessing pipeline included format conversion, VAD (using Silero-VAD), and embedding extraction (using SpeechBrain ECAPA 192-dim embedding).

\subsubsection{Age Regressor}
\label{sssec:age_clf}
\paragraph{Training and evaluating on VoxCeleb2 dataset.}
The age regressor was trained on the VoxCeleb2 multilingual dataset. We used the train-test split described in~\cite{DBLP:journals/corr/abs-2109-13510} maintaining the original test set. The training set was further divided into training and validation sets at the speaker level using a stratified split approach:
\begin{enumerate}
  \item We first created five age bins across the entire age range.
  \item We then combined these age bins with gender information to create stratification groups.
  \item Using these stratification groups, we split the training data into training and validation sets with an 80\%-20\% ratio, ensuring that each set maintained a similar distribution of age groups and genders.
\end{enumerate}

We evaluated three regression models: XGBoost, Support Vector Regression (SVR), and ANN. The Optuna library was employed for hyperparameter tuning, with 200 study trials conducted for each model. We implemented a weighted training approach to address the imbalanced nature of age distribution in the dataset. This involved calculating sample weights inversely proportional to the frequency of age groups (5-year wide bins), ensuring that underrepresented age ranges received appropriate attention during training.

The models were evaluated using MAE and RMSE metrics on the test set. The best-performing model for each feature set was selected based on these metrics.

\paragraph{Training and evaluating on DARPA-TIMIT dataset.}
We extended our approach to the TIMIT dataset for our age regression task. TIMIT provides a predefined train-test split, which we adhered to in our experiments. To create a validation set, we employed the same 80\%-20\% stratified split strategy used with VoxCeleb2 for train-validation division within the original training set.
We applied the same preprocessing and feature extraction methods as in the VoxCeleb2 experiments.

The model architectures, hyperparameter tuning process, and evaluation metrics remained consistent with our VoxCeleb2 methodology. This consistency allowed for a direct comparison of model performance across datasets, providing insights into the generalizability of our approach.

\paragraph{Combined Dataset Training and Cross-Evaluation of Age Regression Models.}
The age distributions in the TIMIT and VoxCeleb2 datasets exhibit notable differences, with TIMIT featuring a more concentrated range of adult speakers. We conducted a cross-evaluation study to assess the generalizability and robustness of our age regression models.

Initial cross-dataset evaluations showed that models trained on individual datasets had limited generalization capabilities. To address this limitation, we developed an integrated approach combining both datasets. The stratified training and validation sets from VoxCeleb2 and TIMIT were merged while maintaining their respective age and gender distributions. We then trained models on this combined dataset, employing Optuna for hyperparameter optimization across 200 trials. We implemented inverse frequency weighting in the training process to address the inherent age distribution imbalance. 

\subsubsection{Height Regressor}
\label{sssec:height_reg}
\paragraph{Training and evaluating on VoxCeleb2 dataset.}
To train the height regression model on the VoxCeleb dataset, we queried the height parameter from Wikidata and converted it to centimeters. The height data of 1715 persons of VoxCeleb1 in conjunction with VoxCeleb2 datasets were found. We are training and testing on VoxCeleb2 alone, which leaves us with 1621 celebrities and their 38425 recordings.

\paragraph{Evaluating on DARPA-TIMIT test set.}
To evaluate the generalization capabilities of our height regression models, we conducted cross-dataset validation using the TIMIT test set. The TIMIT dataset provides height information for all speakers, measured under controlled conditions, unlike the VoxCeleb2 dataset, where heights were obtained from Wikidata.  

\section{Results}
\label{sec:results}
This section presents the evaluation results of our custom-trained models across different datasets, summarizes their performance, and demonstrates the framework's application to movie characters' analysis.

\subsection{Gender Classifier}
\subsubsection{Training and Evaluating on VoxCeleb2 dataset}
The evaluation results of the gender classification model trained on VoxCeleb2 are presented in Table~\ref{Tab:AccGender}, which depicts the accuracy metric on the test set. We also measured the mean inference time for the SpeechBrain (SB) 192 ECAPA embedding for the test set evaluation. This additional performance metric provides insights into the computational efficiency of our models. Table~\ref{Tab:CompF1Score} shows the F1-score of the best-performing model (SVM) in comparison to the best model from the ``VoxCeleb enrichment" research~\cite{DBLP:journals/corr/abs-2109-13510}.
Results revealed that SVM classification achieved 98.1\% and 98.9\% accuracy using SpeechBrain 512-dimensional x-vector embeddings and ECAPA embeddings, respectively. The combination of ECAPA embeddings with Librosa features (223 features) yielded 98.7\% accuracy, which is less than ECAPA alone, indicating that increased dimensionality did not improve results. While the baseline Librosa's 31 acoustic features achieved accuracies between 86.1-88.3\%, embedding-based methods consistently produced accuracies above 90\%. Measured processing times for ECAPA embeddings on the test set ranged from 12.3 ms for Logistic Regression to 1.05 minutes for the SVM.

\footnotetext[7]{\label{abbr}SB - Speech Brain, emb. - embedding}
\begin{table*}[ht]
\centering
\footnotesize
\begin{adjustbox}{width=1\textwidth}
\begin{tabular}{lp{20mm}p{20mm}p{20mm}p{20mm}p{20mm}}
\toprule
\textbf{Classifier} & \textbf{\makecell[l]{SB\footnoteref{abbr} 512 \\ xvect emb.}} & \textbf{\makecell[l]{(1) SB 192 \\ ecapa emb.}} & \textbf{\makecell[l]{Pyannote 512 \\ emb.}} & \textbf{\makecell[l]{(2) Librosa \\ 31 features}} & \textbf{\makecell[l]{(1) \& (2) \\ 223 features}} \\ 
\midrule
Logistic Regression & \textbf{98.1} & 67.4 
$_{(12.3\text{ ms})}$ & 90.4 & 86.1 & 90.0 \\
SVM & \textbf{98.1} & \textbf{98.9}  $_{(1.05\text{ min})}$ & \textbf{98.0} & 88.0 & \textbf{98.7} \\
XGBoost & \textbf{98.1} & 97.6  $_{(741\text{ ms})}$ & 96.9 & 87.8 & 95.8 \\
ANN & 98.0 & 98.6  $_{(1.27\text{ s}})$ & 97.6 & \textbf{88.3} & 98.3 \\ 
\bottomrule
\end{tabular}
\end{adjustbox}
\caption{Accuracy percent of gender classification models on VoxCeleb2 test set}
\label{Tab:AccGender}
\end{table*}

\footnotetext[8]{\label{ds_diff}We followed a similar dataset division as~\cite{DBLP:journals/corr/abs-2109-13510}, but employed a training-validation split rather than 5-fold CV for the training set. We retained the same test-set IDs.}
\begin{table}[ht]
\centering
\footnotesize
\begin{adjustbox}{width=1.0\linewidth}
\begin{tabular}{lc}
\toprule
\textbf{Model} & \textbf{F1-Score\footnoteref{ds_diff}} \\
\midrule
VANPY gender classification model (SVM) & \textbf{0.9885} \\ 
VoxCeleb Enrichment Best Model~\cite{DBLP:journals/corr/abs-2109-13510} & 0.9829 \\
\bottomrule
\end{tabular}
\end{adjustbox}
\caption{Comparison of F1-score performance for gender classification: VANPY SVM vs VoxCeleb Enrichment Best Model}
\label{Tab:CompF1Score}
\end{table}

\subsubsection{Evaluating on Mozilla Common Voice v10.0 English test set}
Results showed lower accuracies of the same model on the Mozilla Common Voice dataset compared to VoxCeleb2. SVM classification maintained the highest performance, reaching 92.3\% with ECAPA embeddings and 92.1\% with the combined ECAPA-Librosa features. %Librosa's acoustic features baseline achieved between 70.3-76.3\% accuracy, while embedding-based methods consistently performed above 87\%. 
Logistic regression showed notably reduced performance compared to other classifiers, particularly with ECAPA embeddings (63.5\%).
Table~\ref{Tab:AccGenderMCV} concludes the evaluation results.

\footnotetext[9]{\label{validated_mcv}Having 'up\_votes' \textgreater  0 and 'down\_votes' = 0}
\begin{table*}[ht]
\centering
\footnotesize
\begin{adjustbox}{width=1\textwidth}
\begin{tabular}{lp{20mm}p{20mm}p{20mm}p{20mm}p{20mm}}
\toprule
\textbf{Classifier} & \textbf{\makecell[l]{SB\footnoteref{abbr} 512 \\ xvect emb.}} & \textbf{\makecell[l]{(1) SB 192 \\ ecapa emb.}} & \textbf{\makecell[l]{Pyannote 512 \\ emb.}} & \textbf{\makecell[l]{(2) Librosa \\ 31 features}} & \textbf{\makecell[l]{(1) \& (2) \\ 223 features}} \\ 
\midrule
Logistic Regression & 89.9 & 63.5 & 79.8 & 70.3 & 77.5 \\
SVM & \textbf{90.6} & \textbf{92.3} & 88.7 & 73.8 & \textbf{92.1} \\
XGBoost & 89.4 & 90.4 & 87.9 & 74.0 & 87.2 \\
ANN & 89.0 & 91.3 & \textbf{89.3} & \textbf{76.3} & 91.3 \\ 
\bottomrule
\end{tabular}
\end{adjustbox}
\caption{Accuracy percent of gender classification models on Mozilla Common Voice v10.0 English validated test set\protect\footnoteref{validated_mcv}}
\label{Tab:AccGenderMCV}
\end{table*}

\subsubsection{Evaluating on DARPA-TIMIT test set}
Table~\ref{Tab:AccGenderTIMIT} summarizes the evaluation results for gender classification trained on VoxCeleb2 on the TIMIT test set. On the TIMIT test set, SVM classification achieved the highest accuracies across all feature types, reaching 99.6\% with ECAPA embeddings and 99.4\% with SpeechBrain x-vector embeddings. Combining ECAPA and Librosa features (99.2\%) did not improve upon ECAPA's individual performance (99.6\%).

\begin{table*}[ht]
\centering
\footnotesize
\begin{adjustbox}{width=1\textwidth}
\begin{tabular}{lp{20mm}p{20mm}p{20mm}p{20mm}p{20mm}}
\toprule
\textbf{Classifier} & \textbf{\makecell[l]{SB\footnoteref{abbr} 512 \\ xvect emb.}} & \textbf{\makecell[l]{(1) SB 192 \\ ecapa emb.}} & \textbf{\makecell[l]{Pyannote 512 \\ emb.}} & \textbf{\makecell[l]{(2) Librosa \\ 31 features}} & \textbf{\makecell[l]{(1) \& (2) \\ 223 features}} \\ 
\midrule
Logistic Regression & 99.2 & 65.9 & 94.6 & 72.3 & 81.1 \\
SVM & \textbf{99.4} & \textbf{99.6} & \textbf{99.2} & \textbf{79.8} & \textbf{99.2} \\
XGBoost & 99.2 & 98.1 & 98.4 & 70.0 & 89.0 \\
ANN & 98.7 & 99.3 & 98.6 & 78.0 & 98.8 \\ 
\bottomrule
\end{tabular}
\end{adjustbox}
\caption{Accuracy percent of gender classification models on TIMIT test set}
\label{Tab:AccGenderTIMIT}
\end{table*}

\subsection{Emotion Recognition Classifier}
\begin{figure}
 \centering
 \captionsetup{justification=centering}
  \centering
  \includegraphics[width=\linewidth]{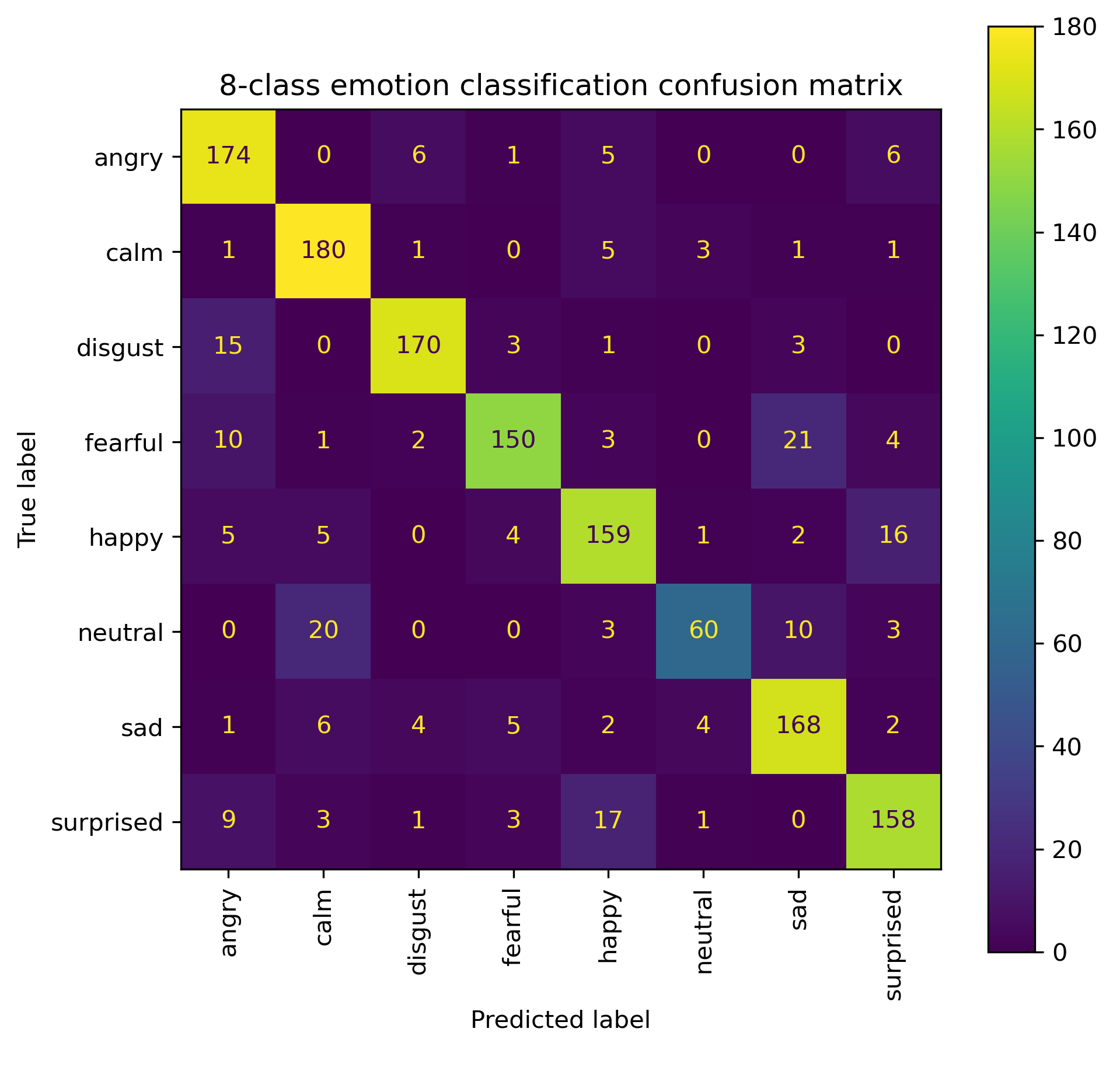}
  \label{fig:sub1}
  \caption{Emotion classification confusion matrix for Eight classes}
 \label{figure:emo_cm8}
\end{figure}

\begin{figure}
  \centering
  \includegraphics[width=\linewidth]{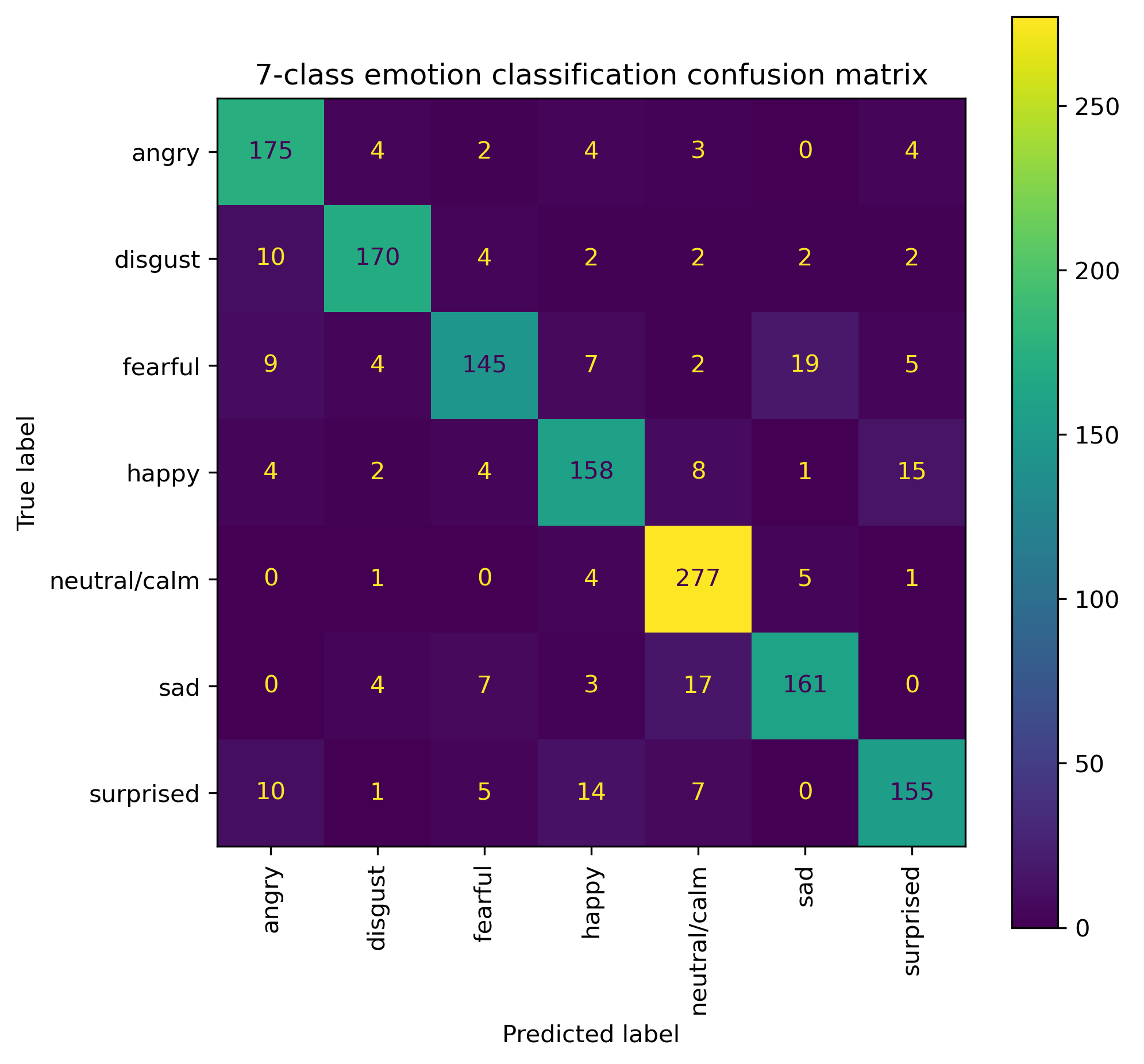}
  \label{fig:sub2}
 \caption{Emotion classification confusion matrix for Seven classes}
 \label{figure:emo_cm7}
\end{figure}

The best-performing SVM model achieved an overall mean accuracy of 84.71\% on the entire dataset using 5-fold cross-validation. Analysis of the most common misclassifications revealed that 'neutral' emotions were frequently misclassified as 'calm', representing one of the primary sources of error. This confusion is understandable, given the subtle distinctions between these two emotional states.
Figures~\ref{figure:emo_cm8},~\ref{figure:emo_cm7} display the confusion matrices for 8-class- and 7-class classifications, respectively, visually representing the model's performance across different emotion categories. Additionally, Table~\ref{Tab:EmotionClassifierComparison} presents our accuracy results compared to previous works in the field.

\begin{table*}[ht]
\centering
\footnotesize
\begin{adjustbox}{width=1\textwidth}
\begin{tabular}{llc}
\toprule
\textbf{Model} & \textbf{Evaluation Method} & \textbf{Accuracy (\%)} \\
\midrule
Sequential bi-LSTM on flatten-810~\cite{s21227665} & 5-fold CV & 57.08 \\
SVM on continuous Wavelet~\cite{7843306} & 5-fold CV & 60.1 \\
Gated ResNet on spectrogram~\cite{zeng2019spectrogram} & 5-fold CV & 64.48 \\
CNN on mel spectrogram~\cite{9640995} & - & 68 \\
CNN on audio features~\cite{ISSA2020101894} & 20\% test set & 71.61 \\
CNN on mel spectrogram and gender~\cite{9640995} & - & 75 \\
CNN-14 on mel spectrograms~\cite{s21227665} & 5-fold CV & 76.58 \\
Dynamic clustering bi-LSTM on spectrogram~\cite{9078789} & 5-fold CV & 77.02 \\
pre-trained xlsr-Wav2Vec2.0 and MLP on raw audio~\cite{app12010327} & 5-fold CV & 81.82 \\
Multimodal Emotion Recognition (audio and video)~\cite{app12010327} & 5-fold CV & 86.7 \\
CNN on mel spectrogram trained on RAVDESS and TESS datasets~\cite{9640995} & - & 89 \\
\midrule
VANPY emotion recognition on ECAPA SpeechBrain embedding & 5-fold CV & 84.71 \\ 
\bottomrule
\end{tabular}
\end{adjustbox}
\caption{Accuracy comparison of emotion classifier models (8 classes) on the RAVDESS dataset.}
\label{Tab:EmotionClassifierComparison}
\end{table*}

\subsection{Age Regressor}

\subsubsection{Training and Evaluating on VoxCeleb2 dataset}

Table~\ref{Tab:MAEAge} presents our age regression models' results trained on VoxCeleb2. Age regression results on the VoxCeleb2 validated test set demonstrated that SVR consistently outperformed other classifiers, achieving the lowest MAE across all feature types (7.88-8.22 years). The combination of ECAPA embeddings with Librosa features resulted in the best SVR performance (MAE = 7.88 years), while Librosa features alone yielded higher errors across all models (MAE $>$ 10.66 years).

\begin{table*}[ht]
\centering
\footnotesize
\begin{adjustbox}{width=1\textwidth}
\begin{tabular}{lp{20mm}p{20mm}p{20mm}p{20mm}p{20mm}}
\toprule
\textbf{Classifier} & \textbf{\makecell[l]{SB\footnoteref{abbr} 512 \\ xvect emb.}} & \textbf{\makecell[l]{(1) SB 192 \\ ecapa emb.}} & \textbf{\makecell[l]{Pyannote 512 \\ emb.}} & \textbf{\makecell[l]{(2) Librosa \\ 31 features}} & \textbf{\makecell[l]{(1) \& (2) \\ 223 features}} \\ 
\midrule
SVR & \textbf{8.22} & \textbf{7.89} & \textbf{8.20} & \textbf{10.66} & \textbf{7.88} \\
XGBoost & 9.13 & 10.20 & 9.11 & 11.29 & 9.94 \\
ANN & 8.51 & 8.64 & 8.79 & 10.74 & 8.37 \\ 
\bottomrule
\end{tabular}
\end{adjustbox}
\caption{MAE of age regression models on VoxCeleb2 test set (in years)}
\label{Tab:MAEAge}
\end{table*}

\subsubsection{Training and Evaluating on DARPA-TIMIT test set}
Our results of the age regression model, which was trained and evaluated on the TIMIT dataset, are presented in Table~\ref{Tab:MAEAgeTIMIT}. Models that used SpeechBrain ECAPA embeddings (192‑dimensional) produced the lowest errors for every classifier; the ANN achieved the overall best MAE of 4.95 years, with SVR and XGBoost close behind. 

\begin{table*}[hp]
\centering
\footnotesize
\begin{adjustbox}{width=1\textwidth}
\begin{tabular}{lp{20mm}p{20mm}p{20mm}p{20mm}p{20mm}}
\toprule
\textbf{Classifier} & \textbf{\makecell[l]{SB\footnoteref{abbr} 512 \\ xvect emb.}} & \textbf{\makecell[l]{(1) SB 192 \\ ecapa emb.}} & \textbf{\makecell[l]{Pyannote 512 \\ emb.}} & \textbf{\makecell[l]{(2) Librosa \\ 31 features}} & \textbf{\makecell[l]{(1) \& (2) \\ 223 features}} \\ 
\midrule
SVR & \textbf{5.39} & 5.19 & 5.19 & 5.84 & 5.22 \\
XGBoost & 8.45 & 5.40 & 8.79 & 12.10 & 8.44 \\
ANN & 5.66 & \textbf{4.95} & \textbf{5.14} & \textbf{5.56} & \textbf{5.09} \\ 
\bottomrule
\end{tabular}
\end{adjustbox}
\caption{MAE of age regression models on TIMIT test set (in years)}
\label{Tab:MAEAgeTIMIT}
\end{table*}

\subsubsection{Combined Dataset Training and Cross-Evaluation of Age Regression Models}
\vspace{5pt}
Cross-dataset evaluation of the models:
\begin{itemize}
    \item The best-performing model trained on VoxCeleb2, when evaluated on the TIMIT test set, yielded a MAE of 13.38 years.
    \item Conversely, the optimal model trained on TIMIT, when applied to the VoxCeleb2 test set, achieved an MAE of 12.19 years.
\end{itemize}

The evaluation results of the combined-dataset approach are presented in 
Table~\ref{Tab:MAEAgeCombined}.
The combined-dataset approach yielded MAEs between single-dataset training and cross-dataset evaluation: using ECAPA embeddings with Librosa features, ANN achieved 8.09, 4.93, and 6.93 years on VoxCeleb2, TIMIT, and combined sets, respectively. This shows considerable improvement over cross-dataset evaluation results ($>$12 years MAE).

\begin{table*}[hp]
\centering
\footnotesize
\begin{adjustbox}{width=1\textwidth}
\begin{tabular}{lp{20mm}p{20mm}p{20mm}p{20mm}}
\toprule
\textbf{Feature Set} & \textbf{Model} & \textbf{VoxCeleb2} & \textbf{TIMIT} & \textbf{Combined} \\ 
\midrule
\multirow{3}{*}{\makecell[l]{(1) SB 192\\ecapa emb.}} 
& SVR & \textbf{8.37} & 5.62 & 7.33 \\
& XGBoost & 9.89 & 5.72 & 8.31 \\
& ANN & 8.40 & \textbf{5.10} & \textbf{7.15} \\
\midrule
\multirow{3}{*}{\makecell[l]{SB 512\\xvect emb.}} 
& SVR & 8.82 & 5.63 & 7.61 \\
& XGBoost & 9.26 & 5.62 & 7.88 \\
& ANN & \textbf{8.25} & \textbf{5.15} & \textbf{7.08} \\
\midrule
\multirow{3}{*}{\makecell[l]{Pyannote 512\\emb.}} 
& SVR & \textbf{8.38} & 5.45 & \textbf{7.27} \\
& XGBoost & 9.91 & 5.62 & 8.28 \\
& ANN & 8.91 & \textbf{5.08} & 7.46 \\
\midrule
\multirow{3}{*}{\makecell[l]{(2) Librosa\\31 features}} 
& SVR & 11.17 & * & 9.11 \\
& XGBoost & 13.06 & \textbf{5.87} & 11.18 \\
& ANN & \textbf{10.87} & * & \textbf{9.06} \\
\midrule
\multirow{3}{*}{\makecell[l]{(1) \& (2)\\223 features}} 
& SVR & 8.68 & 6.00 & 7.69 \\
& XGBoost & 10.10 & 5.37 & 8.37 \\
& ANN & \textbf{8.09} & \textbf{4.93} & \textbf{6.93} \\
\bottomrule
\end{tabular}
\end{adjustbox}
\caption{MAE evaluation results on VoxCeleb2, TIMIT, and Combined test sets (in years). * indicates invalid results ($>$ 100)}
\label{Tab:MAEAgeCombined}
\end{table*}

\subsection{Height Regressor}
\subsubsection{Training and Evaluating on VoxCeleb2 dataset}
Height regression on the VoxCeleb2 test set (see Table~\ref{Tab:MAEHeight}) showed that SVR with ECAPA embeddings achieved the lowest MAE of 6.01 cm, closely followed by combined features (6.10 cm). 
\begin{table*}[ht]
\centering
\footnotesize
\begin{adjustbox}{width=1\textwidth}
\begin{tabular}{lp{20mm}p{20mm}p{20mm}p{20mm}p{20mm}}
\toprule
\textbf{Classifier} & \textbf{\makecell[l]{SB\footnoteref{abbr} 512 \\ xvect emb.}} & \textbf{\makecell[l]{(1) SB 192 \\ ecapa emb.}} & \textbf{\makecell[l]{Pyannote 512 \\ emb.}} & \textbf{\makecell[l]{(2) Librosa \\ 31 features}} & \textbf{\makecell[l]{(1) \& (2) \\ 223 features}} \\ 
\midrule
SVR & 6.26 & \textbf{6.01} & \textbf{6.42} & 9.77 & \textbf{6.10} \\
XGBoost & 6.36 & 6.84 & 6.85 & 7.86 & 6.51 \\
ANN & \textbf{6.24} & 7.74 & 6.63 & \textbf{7.55} & 6.62 \\ 
\bottomrule
\end{tabular}
\end{adjustbox}
\caption{MAE of height regression models on VoxCeleb2 test set (in centimeters)}
\label{Tab:MAEHeight}
\end{table*}

\subsubsection{Evaluating on DARPA-TIMIT test set}
Table~\ref{Tab:MAETIMITHeight} summarizes the evaluation performance of our height regression model, which was trained on VoxCeleb2, when applied to the TIMIT test set. The SVR with ECAPA embeddings achieves the lowest MAE of 6.02 cm, closely matching its performance on VoxCeleb2 (6.01 cm).

\begin{table*}[ht]
\centering
\footnotesize
\begin{adjustbox}{width=1\textwidth}
\begin{tabular}{lp{20mm}p{20mm}p{20mm}p{20mm}p{20mm}}
\toprule
\textbf{Classifier} & \textbf{\makecell[l]{SB\footnoteref{abbr} 512 \\ xvect emb.}} & \textbf{\makecell[l]{(1) SB 192 \\ ecapa emb.}} & \textbf{\makecell[l]{Pyannote 512 \\ emb.}} & \textbf{\makecell[l]{(2) Librosa \\ 31 features}} & \textbf{\makecell[l]{(1) \& (2) \\ 223 features}} \\ 
\midrule
SVR & 6.31 & \textbf{6.02} & 6.09 & 11.79 & \textbf{6.23} \\
XGBoost & \textbf{6.16} & 7.37 & 6.60 & 6.87 & 6.38 \\
ANN & 6.47 & 6.86 & \textbf{6.06} & \textbf{6.39} & 6.39 \\ 
\bottomrule
\end{tabular}
\end{adjustbox}
\caption{MAE of height regression models on TIMIT test set (in centimeters)}
\label{Tab:MAETIMITHeight}
\end{table*}

\subsection{Custom-Trained Models Performance Summary}

The models in our framework, as detailed in Table~\ref{tab:model_performance}, demonstrate robust performance across multiple datasets.

\begin{table}
\caption{Overview of VANPY Models' Performances.}
\centering
\footnotesize
\begin{adjustbox}{width=\linewidth}
\begin{tabular}{llr}
\toprule
\textbf{Model} & \textbf{Dataset} & \textbf{Performance} \\
\midrule
\multirow{3}{*}{Gender Identification (Accuracy)} 
 & VoxCeleb2 & 98.9\% \\
 & Mozilla Common Voice v10.0 & 92.3\% \\
 & TIMIT & 99.6\% \\
\midrule
\multirow{2}{*}{Emotion Recognition (Accuracy)} 
 & RAVDESS (8-class) & 84.71\% \\
 & RAVDESS (7-class) & 86.24\% \\
\midrule
\multirow{3}{*}{Age Estimation (MAE in years)} 
 & VoxCeleb2 & 7.88 \\
 & TIMIT & 4.95 \\
 & Combined VoxCeleb2-TIMIT & 6.93 \\
\midrule
\multirow{2}{*}{Height Estimation (MAE in cm)} 
 & VoxCeleb2 & 6.01 \\
 & TIMIT & 6.02 \\
\bottomrule
\end{tabular}
\end{adjustbox}

\label{tab:model_performance}
\end{table}

\vspace{-5pt}

\subsection{Use-case Example}
\label{sec:usecase}
As a use case, we utilized annotated subtitles from the movie "Pulp Fiction", sourced from the Speaker Naming in Movies dataset \cite{DBLP:journals/corr/abs-1809-08761}. The audio track was partitioned into segments based on these subtitles, refined by extracting the first intense speech segment using the SileroVAD to remove non-vocal data. We selected segments of at least 1 second long and those that were not classified by the YAMNet component as music. This filtering resulted in 906 segments. Subsequently, speaker embedding was executed using SpeechBrain and Librosa feature extractor components, after which several components classified the data: SpeechBrain IEMOCAP emotion classifier, VANPY gender classifier, VANPY age estimator, VANPY height estimator, and Wav2Vec2 arousal, dominance and valence estimator.

\begin{figure}[H]
    \centering
    \begin{minipage}{0.5\textwidth}
        \centering
        \includegraphics[width=0.9\textwidth]{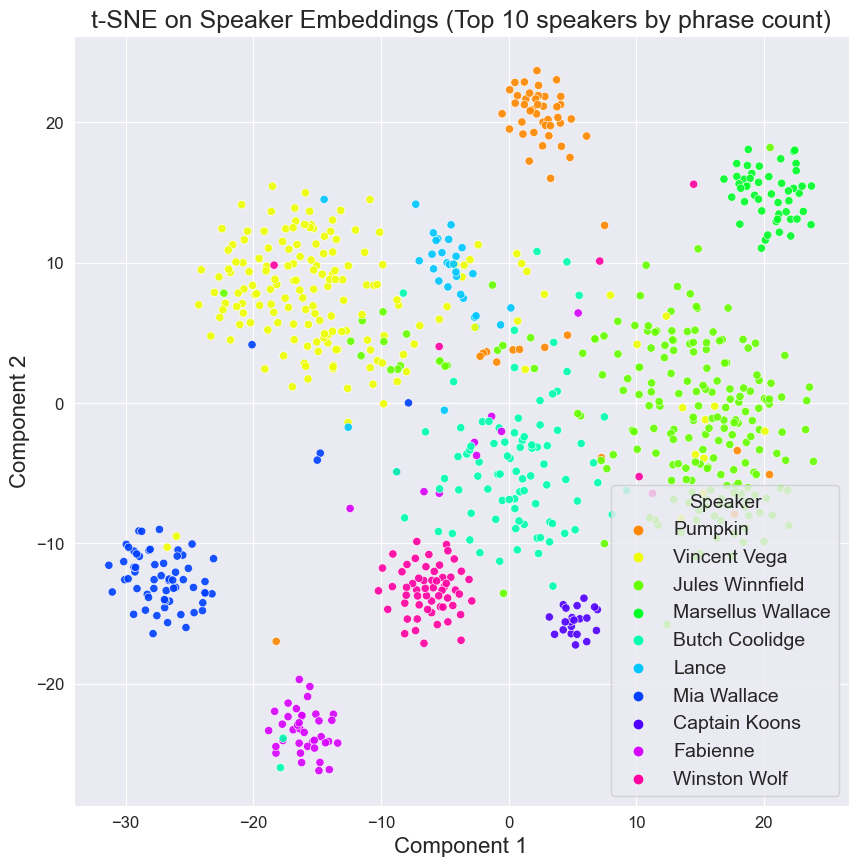}
        \caption{t-SNE representation of speaker embeddings}
        \label{figure:tsne_plot}
    \end{minipage}
\end{figure}

The two-component t-SNE clustering of the speaker embeddings - which were later used by VANPY and emotion classification models - is illustrated in Figure~\ref{figure:tsne_plot}.

For comparison purposes, the classification results were juxtaposed with publicly accessible data about the actors, encompassing gender, age, and height. For each speaker, we took the mean of the classification results. The age of each actor or actress was determined by calculating the difference between their birth year and the film's release year, and this was compared to the average predicted value from the age classifier.

\begin{figure}[H]
         \centering
         \begin{minipage}{0.5\textwidth}
         \includegraphics[width=0.95\linewidth]{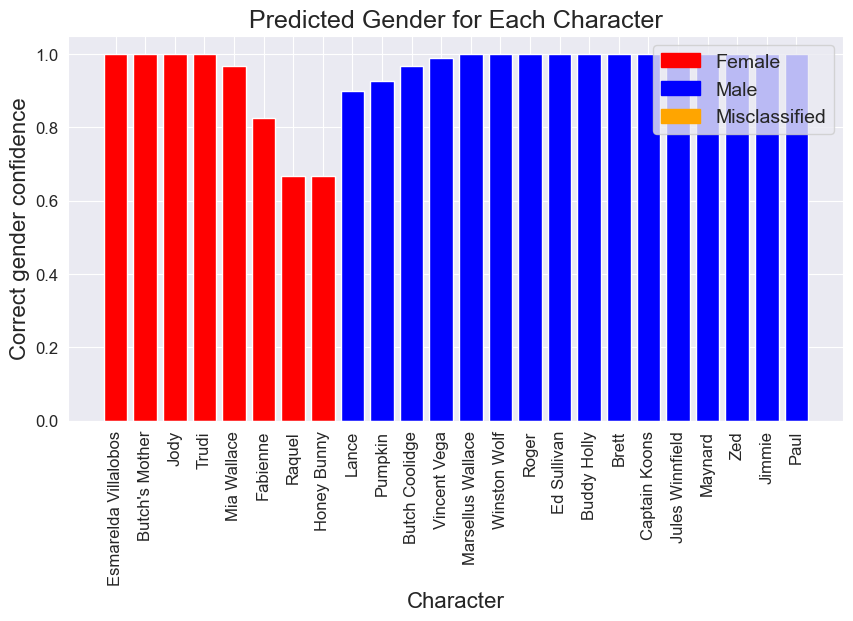}
         \caption{Gender recognition confidence}
         \label{fig:gender-pulp_fiction}
         \end{minipage}
\end{figure}

\begin{figure}[H]
     \centering
         \begin{minipage}{\linewidth}
         \includegraphics[width=1\linewidth]{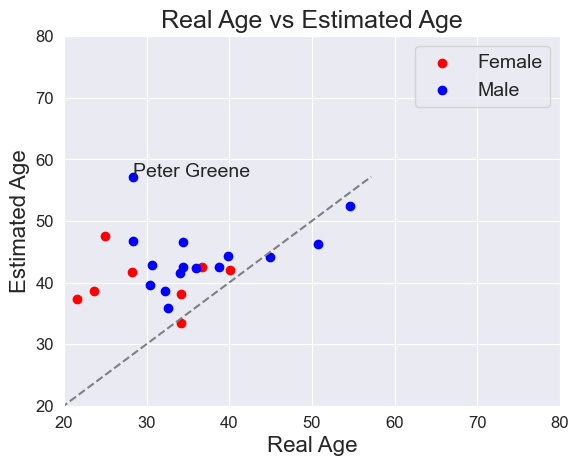}
         \caption{Age estimation}
         \label{fig:age-pulp_fiction}
     \end{minipage}

    \vspace{30pt}

    \begin{minipage}{\linewidth}
        Figure~\ref{fig:emotions-pulp_fiction} exhibits the distribution of expressed emotions by the speaker, normalized by their total utterances. From this visualization, we can speculate about the temper of the characters and see some global tendencies. For example, we can see that most characters express themselves in a calm and neutral manner, with the exception of a relatively high number of angry remarks.
    \end{minipage}
    
\end{figure}

Figure~\ref{fig:gender-pulp_fiction} shows that no speakers were misclassified by gender and that the overall confidence level exceeds 95\%. Figure~\ref{fig:age-pulp_fiction} compares the actors' ages at the time of filming against the age estimates provided by the model. These estimations tend to overstate the age of both male and female actors, with one significant outlier being Peter Greene, who was classified on two speech segments with non-neglectable background noise.

\begin{figure*}[hp]
    \centering
    
    \begin{minipage}{1\textwidth}
         \includegraphics[width=1\textwidth]{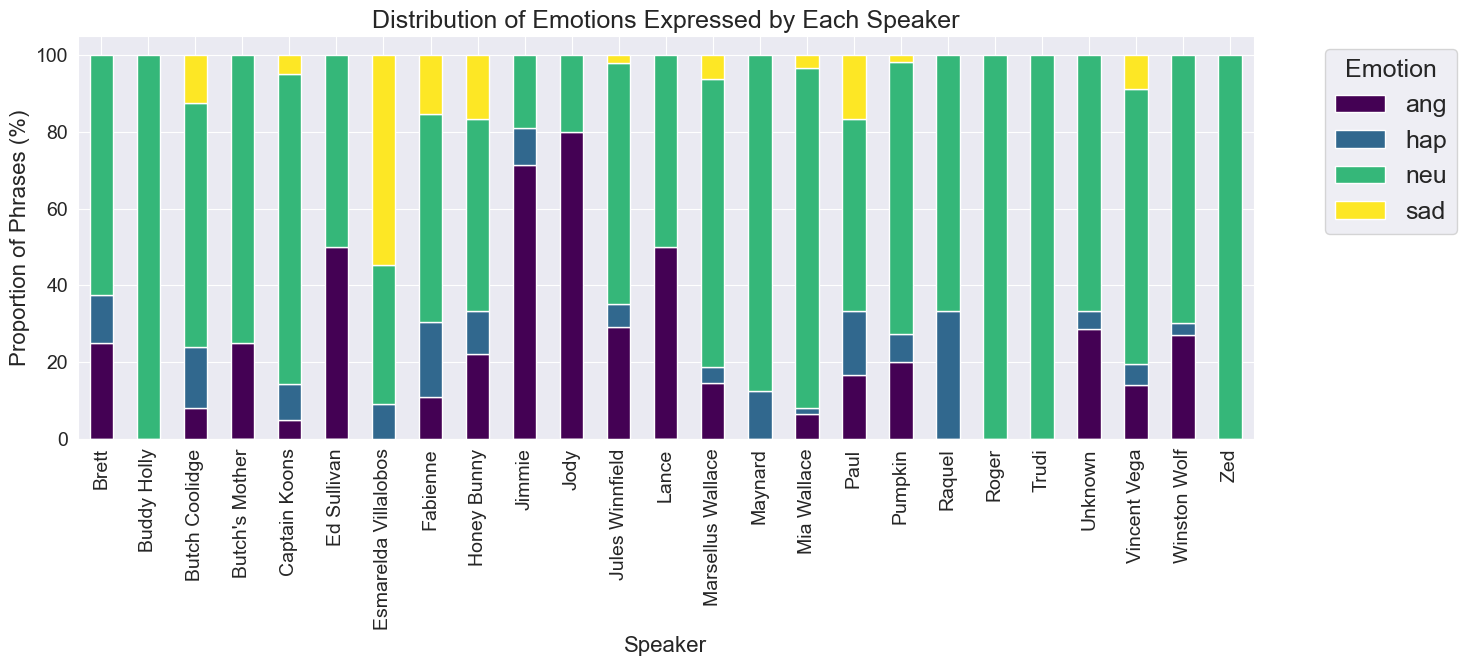}
        \caption{Expressed emotion normalized by speaker utterances}
        \label{fig:emotions-pulp_fiction}
     \end{minipage}
    
\end{figure*}

\begin{figure}[H]
    \centering
    \begin{minipage}{0.6\textwidth}
        \includegraphics[width=0.9\textwidth]{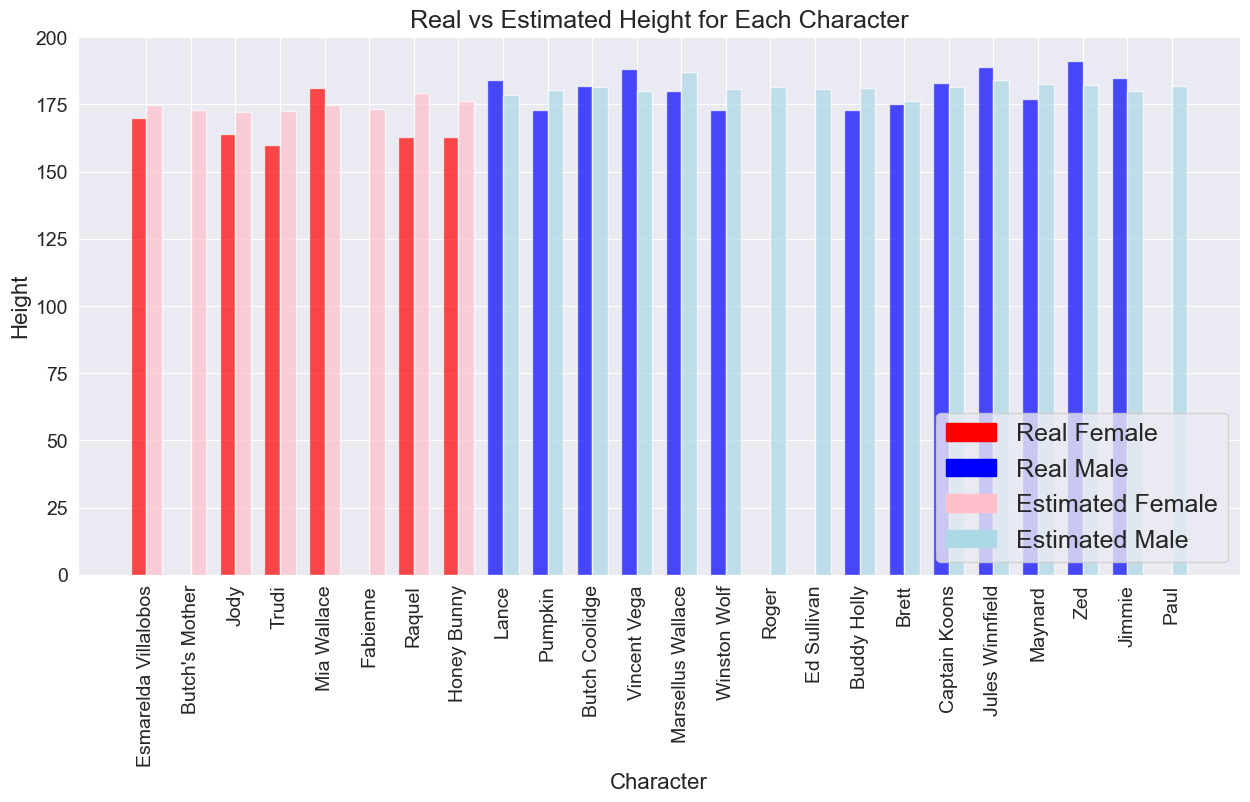}
        \captionsetup{justification=centering,margin=0.5cm}
        \caption{Height estimation}
        \label{fig:height-pulp_fiction}
    \end{minipage}

    \vspace{30pt}

     \begin{minipage}{0.5\textwidth}
        Figure~\ref{fig:height-pulp_fiction} illustrates the height estimation for each character. The model's estimates tended to be higher than the actual heights for female characters, resulting in MAE of 5.5 cm for men and 10.2 cm for women.
         
     \end{minipage}
\end{figure}

\begin{figure}[H]
 \centering
 \captionsetup{justification=centering}
 \includegraphics[width=\linewidth]{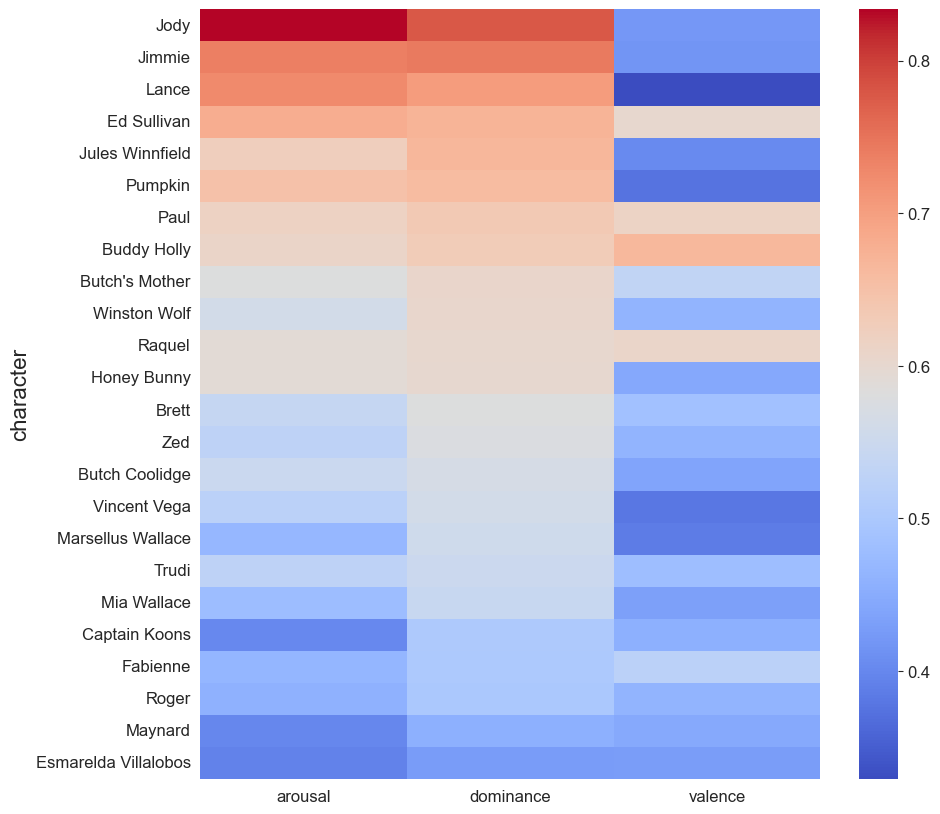}
  \caption{Arousal, Dominance, and Valence mean estimations by speaker}
	\label{figure:adv_plot}
\end{figure}

Figure~\ref{figure:adv_plot} demonstrates each character's mean arousal (which describes emotional intensity), dominance, and valence characteristics (corresponding to positive emotions).

\section{Discussion}
\label{sec:discussion}
We have introduced VANPY, an end-to-end framework for speaker characterization, detailing its architecture and pre-integrated components. We developed and evaluated four speaker characterization models: gender and emotion classifiers, along with age and height regression models. We've also showcased the framework's usage by applying it to the "Pulp Fiction" audio track and performing basic analysis on the retrieved speaker classifications.

Regarding the Gender Classifier, the ANN model achieved slightly lower performance on the VoxCeleb2 test set (98.6\% accuracy) compared to the SVM (98.9\% accuracy) when using the Speech Brain ECAPA embedding. However, the ANN's inference time is significantly lower, 49 times faster than the SVM, even on hardware without GPU (1.27 seconds for ANN vs. 63 seconds for SVM, processing 39,246 records). Despite slight differences in dataset splits, our VANPY gender classification model (SVM) achieved an F1-score of 0.9885, outperforming the 0.9829 reported by the VoxCeleb Enrichment Best Model~\cite{DBLP:journals/corr/abs-2109-13510} on the same test set. Our results are competitive with recent state-of-the-art approaches,  such as the wav2vec 2.0 model by \citet{burkhardt2023speechbasedagegenderprediction} (97.8\% on VoxCeleb2) and the pre-trained models of \citet{lastow2022language} (F1-scores of 0.960-0.962 on Mozilla Common Voice). While some models achieve higher accuracy on specific datasets (e.g., \citet{burkhardt2023speechbasedagegenderprediction} achieved 100\% accuracy on TIMIT using a complex DNN architecture with 317.5M parameters and fine-tuned Transformer), our approach demonstrates robust performance across all datasets, with the SVM model using ECAPA embedding consistently emerging as the leading solution, with accuracy ranging from 92.3\% to 99.6\%. Notably, we observed that adding additional features to the embedding did not improve performance in the gender classification task.

For the Emotion Recognition Classifier, the 8-class model achieved an accuracy of 84.71\%, surpassing many existing studies, though not claiming to be state-of-the-art, which currently stands at 89\%. By merging the 'neutral' and 'calm' target classes, the model's performance improved by approximately 1.5\%, yielding an accuracy of 86.24\% in a 5-fold cross-validation. Cross-dataset evaluation remains challenging due to inconsistent emotion labeling schemes and varying numbers of emotion classes across datasets, making direct performance comparisons difficult.

We developed three separate training approaches for the age regression task: single-dataset training using VoxCeleb2, single-dataset training using TIMIT, and a combined-dataset approach. This strategy allowed us to evaluate the impact of dataset size and diversity on model performance. We implemented sample weighting based on age distribution to address the inherent imbalance in both datasets. Additionally, we split the training and validation sets in a stratified manner by dividing samples into groups of gender-age pairs. Interestingly, we observed that SVR consistently outperformed other models when trained on VoxCeleb2 alone, while ANN showed superior performance in both TIMIT-only and combined training scenarios. Our results are comparable to recent state-of-the-art approaches: while \citet{lastow2022language} achieved a better MAE of 4.22 years using the fine-tuned WavLM model, and \citet{burkhardt2023speechbasedagegenderprediction} reported 7.1 years MAE on TIMIT and 7.7 years on VoxCeleb2, our model achieved 7.88 years MAE on VoxCeleb2 (an improvement over the 9.443 years MAE of \citet{DBLP:journals/corr/abs-2109-13510}), and 4.95 years MAE on TIMIT, outperforming \citet{kaushik2021end}'s gender-specific results on TIMIT (5.62 and 6.08 years MAE for male and female speakers, respectively). Our model shows competitive performance across different datasets.

The Height Regressor SVR model with SpeechBrain ECAPA embeddings achieved the lowest MAE (6.02 on the TIMIT test set and 6.01 on the VoxCeleb2 test set). Despite the different data collection methodologies, our models demonstrated robust performance across datasets. We know that \citet{kaushik2021end} achieved a slightly better result (MAE of 5.24 cm for males and 5.09 cm for females) on the same dataset with Multi-Task Learning as described in Section~\ref{ssec:model_inf}.

In our showcase example, the analysis provides insightful perspectives into character characteristics. The gender classifier achieved 100\% accuracy. The height regression model showed a gender-based performance gap, with better accuracy for men (MAE of 5.5 cm) than women (MAE of 10.2 cm). The age estimation exhibited bias toward higher predictions for younger actors, possibly due to movie background noise not represented in the training datasets. While emotion classification requires expert validation, our analysis of the character named Jody revealed consistent emotional intensity, with 4 out of 5 segments containing shouting, supported by high arousal and dominance values in both emotion intensity plots and ``angry'' emotion classification.

\section{Conclusions}
\label{sec:conclusions}
We introduced an open-source end-to-end voice characterization framework. The framework consists of three optional pipelines with numerous implemented components and is easily extendable with additional ones. The performance of the models within VANPY has shown promising results. For instance, the gender identification model achieved an accuracy of 92.3\%-99.6\% depending on the dataset, and the emotion classification model yielded an accuracy of 84.71\% across eight classes. These results demonstrate the robustness and applicability of the framework.

Another contribution of this work is an enhancement for the VoxCeleb datasets, adding a height parameter taken from Wikidata for 1,715 celebrities. This allowed for the development of a height regressor.

In future work, we plan to extend the framework with more pre-trained components, including accent detection, sound quality estimation, and emotion intensity, and even expand to new audio analysis fields, such as music genre and audio classification.

\section*{Code Availability}
\label{sec:code}
The framework is published under the Apache 2.0 license and may be used in different fields and applications (\url{https://github.com/griko/vanpy}). All the models developed in this work are also published as standalone models in the Hugging Face Hub (\url{https://huggingface.co/griko}) and can be used independently from the framework.

\section*{Acknowledgments}
We acknowledge the VoxCeleb2~\cite{Chung18b} and TIMIT~\cite{discacoustic} datasets for providing the training data used in this research. We also extend our gratitude to the SpeechBrain~\cite{speechbrain, Dawalatabad_2021} team for their excellent open-source speech processing toolkit. We also appreciate the contributions of the Librosa~\cite{brian_mcfee_2022_6097378}, Optuna~\cite{akiba2019optunanextgenerationhyperparameteroptimization}, TensorFlow~\cite{tensorflow2015whitepaper}, and scikit-learn~\cite{scikit-learn} teams for developing essential libraries that facilitated feature extraction, model training, and evaluation in this work.

Additionally, we acknowledge the assistance of generative large language models~\cite{OpenAI2025, Anthropic2023}, which were used to enhance the clarity and richness of the text in this paper.

%----------------------------------------------------------------------------------------
%	REFERENCE LIST
%----------------------------------------------------------------------------------------
\phantomsection
\Urlmuskip=0mu plus 1mu\relax
\bibliography{sample}

%----------------------------------------------------------------------------------------

\end{document}